\documentclass[aps]{revtex4}
\usepackage{amsfonts}
\usepackage{graphics}
\usepackage{graphicx}
\usepackage{epsf}
\begin{document}
\title{Supersonic Velocities in Noncommutative Acoustic Black Holes}
\author{M. A. Anacleto, F.A. Brito, E. Passos}
\email{fabrito, anacleto, passos@df.ufcg.edu.br}
\affiliation{Departamento de F\'{\i}sica, Universidade Federal de Campina Grande, 58109-970, Campina Grande, Para\'{\i}ba, Brazil}
\begin{abstract}
In this paper we derive Schwarzschild and Kerr - like noncommutative acoustic black hole metrics in the (3+1)-dimensional noncommutative Abelian Higgs model. We have found that the changing in the Hawking temperature $T_H$ due to spacetime noncommutativity  account for supersonic velocities and analogously may also account for superluminal particles in the general form $(v-c)/{c}={\Delta T_H}/8T_H$. Assuming this form is also valid for gravitational black holes and particle physics, we have found that for $\Delta T_H$ about the heaviest muon neutrino mass and $T_H$ about proton mass gives  $2.12\times 10^{-5}$ which agrees with the recent OPERA experiments. We also consider the Hawking temperature in terms of surface gravity to address the issue of background influence in the neutrino superluminality to show that  our investigations also agree with supernova SN1987A measurements. 
\end{abstract}
\maketitle
\pretolerance10000
\section{Introduction}

 Acoustic black holes possess many of the fundamental properties of the black holes of general relativity and have been extensively studied in the 
literature \cite{MV,Volovik,Unruh}. The connection between black hole physics and the theory of supersonic acoustic flow was established in 1981 by Unruh \cite{Unruh} and has been developed to investigate the Hawking radiation 
and other phenomena for understanding quantum gravity. Hawking radiation is an important quantum effect of black hole physics. In 1974, Hawking combining Einstein's General Relativity and
Quantum Mechanics announced that classically a black hole does not radiate, but when we consider quantum effects emits thermal radiation at a temperature proportional to the horizon surface gravity.

Since the Hawking radiation showed by Unruh \cite{Unruh} is a purely
kinematic effect of quantum field theory, we can study the Hawking radiation process in completely
different physical systems. For example, acoustic horizons are regions where a moving fluid exceeds the local sound speed through 
a spherical surface and possesses many of the properties associated with
the event horizons of general relativity. In particular, the acoustic Hawking radiation when quantized appears as a flux of thermal phonons
emitted from the horizon at a temperature proportional to its surface gravity. Many fluid systems
have been investigated on a variety of analog models of acoustic black holes, including gravity wave~\cite{RS}, water
\cite{Mathis}, slow light \cite{UL}, optical fiber \cite{Philbin} and  electromagnetic waveguide \cite{RSch}. The models of superfluid helium II \cite{Novello}, atomic Bose-Einstein condensates (BEC) \cite{Garay,OL} and one-dimensional Fermi degenerate
noninteracting gas \cite{SG} have been proposed to create an acoustic black hole geometry in
the laboratory.

The purpose of this paper is to investigate the relativistic version of acoustic black holes from the noncommutative Abelian Higgs model. 
Various gravitational black hole solutions of noncommutative spacetime have been found in~\cite{NCBH}, thermodynamic properties of the noncommutative
black hole were studied in~\cite{TPNCBH}, the evaporation of the noncommutative black hole was studied in~\cite{ENCBH}, quantum tunneling of noncommutative Kerr black hole was studied in~\cite{TKerrBH}, quantized entropy was studied in~\cite{entropy}. 

A relativistic version of acoustic black holes has been presented in \cite{Xian,ABP} (see also \cite{Bilic}). This is also motivated by the fact that in high energy physics both strong spacetime noncommutativity and quark gluon plasma (QGP) may take place together. Thus, it seems to be natural to look for acoustic black holes in a QGP fluid with spacetime noncommutativity  in this regime. Acoustic phenomena in QGP matter can be seen in Ref.~\cite{shk} and acoustic black holes in a plasma fluid can be found in Ref.~\cite{BH-plasma}.

Differently of the most cases studied, we consider the acoustic black hole metrics obtained from a relativistic fluid in a noncommutative spacetime. 
 The effects of this
set up is such that the fluctuations of the fluids are also affected.  The sound waves inherit spacetime noncommutativity of the fluid and may lose the Lorentz invariance. As a consequence the Hawking temperature is directly affected by the spacetime noncommutativity. Analogously to Lorentz-violating gravitational black holes \cite{syb,adam}, the effective Hawking temperature of the noncommutativity  acoustic black holes 
now is {\it not} universal for all species of particles. It depends on the maximal attainable velocity of this species.
Furthermore, the acoustic black hole metric can be identified with 
an acoustic Kerr-like black hole. The spacetime noncommutativity  affects the rate of loss of mass of the black hole.  Thus for suitable values of
the spacetime noncommutativity parameter a wider or narrower spectrum of particle wave function can be scattered with increased amplitude by the acoustic black hole. This
increases or decreases the superressonance phenomenon previously studied in \cite{Basak:2002aw,SBP}.

The paper is organized as follows. In Sec.~\ref{II} we obtain a general acoustic black hole metric in the noncommutative  Abelian Higgs model that reveals to be similar to Lorentz violating acoustic black holes \cite{ABP,ABP2} of Lorentz violating Abelian Higgs model \cite{Bazeia:2005tb}. In Sec.~\ref{dp} we address the issue of group velocity, which shows a deviation on the maximal attainable particle velocity on the fluid. The magnitude of the deviation is consistent with that found in Lorentz violating models \cite{codata,Casana:2011bv}. In Sec.~\ref{acoustic-BH-3} we find explicitly Schwarzschild and Kerr - like noncommutative acoustic black holes and address the issue of supersonic (and superluminal) velocities in terms of the variation of the Hawking temperature as a response to the  spacetime noncommutativity. Such a response is given as 
\begin{equation} 
\frac{\Delta T_H}{T_H}=4\theta_3B_3.
\end{equation}
As we shall see this allows us to write the deviation on the maximal attainable particle velocity as simply given by $(v_g-c_s)/{c_s}={\Delta T_H}/8T_H$. We shall also interpret this in particle physics by assuming that 
$\Delta T_H$ is associated with the rest mass of an emitted particle in a {\it black hole decay due to spacetime noncommutativity}, where we assume  $T_H$ about the proton mass. In doing this we achieve a result  consistent  with OPERA experiments \cite{:2011zb} --- see also very recent related theoretical discussions \cite{OPERA2}. In Sec.~\ref{conclu} we make our final conclusions.

\section{The Acoustic Metric in Noncommutative Abelian Higgs Model}
\label{II}

In this section we make an extension of the Abelian Higgs model by modifying its scalar and gauge sector with the Moyal product  \cite{SW,SGhosh,rivelles,revnc}. Thus, 
the Lagrangian of the noncommutative Abelian Higgs model in flat space is
\begin{eqnarray}
\hat{\cal L}&=&-\frac{1}{4}\hat{F}_{\mu\nu}\ast\hat{F}^{\mu\nu} 
+(D_{\mu}\hat{\phi})^{\dagger}\ast D^{\mu}\hat{\phi}+ m^2\hat{\phi}^{\dagger}\ast\hat{\phi}-b\phi^{\dagger}\ast\phi\ast\phi^{\dagger}\ast\phi.
\end{eqnarray}
Now we use the Seiberg-Witten map \cite{SW}, up to the lowest order in the spacetime noncommutative parameter $\theta^{\mu\nu}$, given by 
\begin{eqnarray}
&&\hat{A}_{\mu}=A_{\mu}+\theta^{\nu\rho}A_{\rho}(\partial_{\nu}A_{\mu}-\frac{1}{2}\partial_{\mu}A_{\nu}),
\nonumber\\
&&\hat{F}_{\mu\nu}=F_{\mu\nu}+\theta^{\rho\beta}(F_{\mu\rho}F_{\nu\beta}+A_{\rho}\partial_{\beta}F_{\mu\nu}),
\nonumber\\
&&\hat{\phi}=\phi-\frac{1}{2}\theta^{\mu\nu}A_{\mu}\partial_{\nu}\phi.
\end{eqnarray}
This very useful map leads to the corresponding theory in a commutative spacetime~\cite{SGhosh}
\begin{eqnarray}
\label{acao}
\hat{\cal L}&=&-\frac{1}{4}F_{\mu\nu}F^{\mu\nu}\left(1+\frac{1}{2}\theta^{\alpha\beta}F_{\alpha\beta}\right) 
+\left(1-\frac{1}{4}\theta^{\alpha\beta}F_{\alpha\beta}\right)\left(|D_{\mu}\phi|^2+ m^2|\phi|^2-b|\phi|^4\right)
\nonumber\\
&+&\frac{1}{2}\theta^{\alpha\beta}F_{\alpha\mu}\left[(D_{\beta}\phi)^{\dagger}D^{\mu}\phi+(D^{\mu}\phi)^{\dagger}D_{\beta}\phi \right],
\end{eqnarray}
where $F_{\mu\nu}=\partial_{\mu}A_{\nu}-\partial_{\nu}A_{\mu}$ and  $D_{\mu}\phi=\partial_{\mu}\phi - ieA_{\mu}\phi$. As one knows the parameter $\theta^{\alpha\beta}$ is a constant, real-valued antisymmetric $D\times D$- matrix in $D$-dimensional spacetime with dimensions of length squared. For a review see \cite{revnc}.

Now, in order to find {\it the noncommutative acoustic black hole metric}, let us use the decomposition $\phi=\sqrt{\rho(x,t)}\exp{(iS(x,t))}$ in the original Lagrangian to obtain
\begin{eqnarray}\label{Lag2}
{\cal L}&=&-\frac{1}{4}F_{\mu\nu}F^{\mu\nu}\left(1-2\vec{\theta}\cdot\vec{B}\right)+\tilde{\theta}\left[\partial_{\mu}S\partial^{\mu}S
-2eA_{\mu}\partial^{\mu}S + e^2A_{\mu}A^{\mu} + m^2\right]\rho-\tilde{\theta}b\rho^2
\nonumber\\
&+&\Theta^{\mu\nu}\left[\partial_{\mu}S\partial_{\nu}S-eA_{\mu}\partial_{\nu}S-eA_{\nu}\partial_{\mu}S+e^2A_{\mu}A_{\nu}\right]\rho
+\frac{\rho}{\sqrt{\rho}}\left[\tilde{\theta}\partial_{\mu}\partial^{\mu}+\Theta^{\mu\nu}\partial_{\mu}\partial_{\nu}\right]\sqrt{\rho},
\end{eqnarray}
where $\tilde{\theta}=(1+\vec{\theta}\cdot\vec{B})$, $\vec{B}=\nabla\times\vec{A}$ and $\Theta^{\mu\nu}=\theta^{\alpha\mu}{F_{\alpha}}^{\nu}$.  In our calculations we shall consider the case where there is no 
noncommutativity between spatial and temporal coordinates, that is $\theta^{0i}=0$,  but $\theta^{ij}=\varepsilon^{ijk}\theta^{k}$, $F^{i0}=E^{i}$ and $F^{ij}=\varepsilon^{ijk}B^{k}$.

By using the Lagrangian (\ref{Lag2}) we can find the following equations of motion
\begin{eqnarray}
&&-\partial_{\mu}\left[\tilde{\theta}\rho(\partial^{\mu}S-eA^{\mu})
+\frac{\rho}{2}(\Theta^{\mu\nu}+\Theta^{\nu\mu})(\partial_{\nu}S-eA_{\nu})\right]=0,
\end{eqnarray}
that is
\begin{eqnarray}
\label{cont}
&&-\partial_{t}\left[\tilde{\theta}\rho(\dot{S}-eA_{t})+\frac{\rho\Theta^{jt}}{2}(\partial_{j}S-eA_{j})\right]
\nonumber\\
&&+\partial_{i}\left[-\rho\tilde{\theta}(\partial^{i}S-eA^{i})-\frac{\rho\Theta^{it}}{2}(\dot{S}-eA_{t})
-\frac{\rho}{2}(\Theta^{il}+\Theta^{li})(\partial_{l}S-eA_{l})\right]=0,
\end{eqnarray}
and
\begin{eqnarray}
&&\frac{(\tilde{\theta}\partial_{\mu}\partial^{\mu}+\Theta^{\mu\nu}\partial_{\mu}\partial_{\nu})\sqrt{\rho}}{\sqrt{\rho}}
+\tilde{\theta}\left(\partial_{\mu}S-eA_{\mu}\right)^2
+\Theta^{\mu\nu}\left(\partial_{\mu}S-eA_{\mu}\right)\left(\partial_{\nu}S-eA_{\nu}\right)
+\tilde{\theta}m^2-2\tilde{\theta}b\rho=0,
\end{eqnarray}
that can also be given as
\begin{eqnarray}
\label{fluid}
&&\frac{(\tilde{\theta}\partial_{\mu}\partial^{\mu}+\Theta^{\mu\nu}\partial_{\mu}\partial_{\nu})\sqrt{\rho}}{\sqrt{\rho}}
+\tilde{\theta}\left(\dot{S}-eA_{t}\right)^2
+\tilde{\theta}\left(\partial_{i}S-eA_{i}\right)\left(\partial^{i}S-eA^{i}\right)
\nonumber\\
&&+\Theta^{jt}\left(\partial_{j}S-eA_{j}\right)\left(\dot{S}-eA_{t}\right)
+\Theta^{jl}\left(\partial_{j}S-eA_{j}\right)\left(\partial_{l}S-eA_{l}\right)
+\tilde{\theta}m^2-2\tilde{\theta}b\rho=0.
\end{eqnarray}
For the gauge field $A_{\mu}$, we obtain the modified Maxwell's equations
\begin{eqnarray}
\!\!\!\!\!\!\!&&\partial_{\mu}F^{\mu\nu}+ \frac{1}{4}\partial_{\mu}\left(\theta^{\mu\nu}F_{\alpha\beta}F^{\alpha\beta}
+2\theta^{\alpha\beta}F_{\alpha\beta}F^{\mu\nu}\right)-\frac{1}{2}\theta^{\mu\nu}\partial_{\mu}(u_{\alpha}u^{\alpha}\rho+m^2\rho-b\rho^2)
\nonumber\\
&&-\partial_{\mu}\left[u_{\beta}(\theta^{\mu\beta}u^{\nu}-\theta^{\nu\beta}u^{\mu})\rho\right]
+\partial_{\mu}\left[\frac{\rho}{\sqrt{\rho}}\left(\frac{1}{2}\theta^{\mu\nu}\partial_{\alpha}\partial^{\alpha}
-\theta^{\mu\beta}\partial_{\beta}\partial^{\nu}+\theta^{\nu\beta}\partial_{\beta}\partial^{\mu}\right)\sqrt{\rho}\right]
=2e\rho(1+\vec{\theta}\cdot\vec{B}){u}^{\nu}
\nonumber\\
&&+e\rho {u}^{\mu}(\theta^{\alpha\nu}F_{\alpha\mu}+\theta_{\alpha\mu}F^{\alpha\nu}),
\end{eqnarray}
where we have defined $u^{\mu}=\partial^{\mu}S-eA^{\mu}=(-w,-u^{i})$.  

That is, there exist changes in the Gauss and Amp\`ere laws
\begin{eqnarray}
\nabla\cdot[(1-2\vec{\theta}\cdot\vec{B})\vec{E}]-\theta^{ij}\partial_{i}(u_{j}w\rho)
+\partial_{i}\left[\frac{\rho}{\sqrt{\rho}}\left(\theta^{ij}\partial_{j}\partial^{0}\right)\sqrt{\rho}\right]
=2e\rho(1+\vec{\theta}\cdot\vec{B}){w}+e\rho(\vec{\theta}\times\vec{E})\cdot\vec{u},
\end{eqnarray}
and
\begin{eqnarray}
&&(\nabla\times\vec{B})^{j}-\partial_{t}E^{j}-\frac{1}{4}\theta^{ij}\partial_{i}(F_{\mu\nu}F^{\mu\nu})+2\partial_{t}[(\vec{\theta}\cdot\vec{B})E^{j}]
+2\partial_{i}[(\vec{\theta}\cdot\vec{B})F^{ij}]
+\frac{1}{2}\theta^{ij}\partial_{i}(u_{\alpha}u^{\alpha}\rho+m^2\rho-b\rho^2)
\nonumber\\
&&+\theta^{il}\partial_{i}(u_{l}u^{j}\rho)-\theta^{jl}\partial_{\mu}(u_{l}u^{\mu}\rho)-
\partial_{\mu}\left[\frac{\rho}{\sqrt{\rho}}\left(\frac{1}{2}\theta^{\mu j}\partial_{\alpha}\partial^{\alpha}
-\theta^{\mu\beta}\partial_{\beta}\partial^{j}+\theta^{j\beta}\partial_{\beta}\partial^{\mu}\right)\sqrt{\rho}\right]
\nonumber\\
&&=2e\rho(1+\vec{\theta}\cdot\vec{B})u^{j}+e\rho {w}(\vec{\theta}\times\vec{E})^{j}
+2e\rho(\vec{\theta}\cdot\vec{B})u^{j}
-e\rho(\vec{\theta}\cdot\vec{u})B^{j}-e\rho(\vec{u}\cdot\vec{B})\theta^{j},
\end{eqnarray}
We shall consider plane wave solutions or background fields satisfy the gauge field equations \cite{rivelles}. This allows us to write our acoustic black hole metric in terms of a constant parameter $\vec{\theta}\cdot\vec{B}$ with some freedom to choose it arbitrarily small (or large) depending on the regime where the spacetime noncommutativity takes place --- such a regime is well assumed to happen in the UV regime where very small distance around $\sqrt{\theta}$ can be probed.

The Eq.~(\ref{cont}) is the continuity equation and Eq.~(\ref{fluid}) is an equation describing a hydrodynamical fluid with a term 
$\frac{(\tilde{\theta}\partial_{\mu}\partial^{\mu}+\Theta^{\mu\nu}\partial_{\mu}\partial_{\nu})\sqrt{\rho}}{\sqrt{\rho}}$ called quantum potential (quantum correction term)
\cite{MV}, which can be negligible in the hydrodynamic region. Now we consider perturbations around the background $(\rho_{0}, S_{0})$, with $\rho=\rho_{0}+ \rho_{1}$ and $S=S_{0}+S_{1}$, so we can rewrite (\ref{cont}) and (\ref{fluid}) as
\begin{eqnarray}
&&-\partial_{t}\left[\rho_{0}\left(\tilde{\theta}\dot{S}_{1}+\frac{\Theta^{jt}}{2}\partial_{j}S_{1}\right)
-\rho_{1}\left(\tilde{\theta}w_{0}-\frac{\Theta^{jt}}{2}v^{j}_{0}\right)\right]
\nonumber\\
&&-\partial_{i}\left[\rho_{0}\left(\tilde{\theta}\partial^{i}S_{1}+\frac{\Theta^{it}}{2}\dot{S}_{1}
+\frac{1}{2}(\Theta^{il}+\Theta^{li})\partial_{l}S_{1}\right)
+\rho_{1}\left(-\tilde{\theta}v^{i}_{0}-\frac{\Theta^{it}}{2}w_{t}+\frac{1}{2}(\Theta^{il}+\Theta^{il})v^{l}_{0}\right)\right]=0,
\end{eqnarray}
and
\begin{equation}
-2\tilde{\theta}w_{0}\dot{S}_{1}
-2\tilde{\theta}v^{i}_{0}\partial_{i}S_{1}+\Theta^{lt}(v^{l}_{0}\dot{S_{1}}-w_{0}\partial_{l}S_{1})
+\Theta^{lj}(v^{l}_{0}\partial_{j}S_{1}+v^{j}_{0}\partial_{l}S_{1})- \tilde{\theta}b\rho_{1}=0,
\end{equation}
where we have defined, $w_{0}=-\dot{S}_{0}+eA_{t}$ and $\vec{v}_{0}=\nabla S_{0}+e\vec{A}$ (the local velocity field). 
Thus, the wave equation for the perturbations $S_1$ around the background $S_0$ becomes
\begin{eqnarray}
\label{eqwave}
\partial_{t}[a^{tt}\dot{S}+a^{tj}\partial_{j}S]+\partial_{i}[a^{it}\dot{S}+a^{ij}\partial_{j}S]=0,
\end{eqnarray}
where
\begin{eqnarray}
a^{tt}&=&-\tilde{\theta}\rho_{0}-\frac{2}{b}\left(\tilde{\theta}w^2_{0}-\Theta^{jt}v^{j}_{0}w_{0}\right),
\\
a^{tj}&=&-\frac{1}{2}\rho_{0}\Theta^{jt}
-\frac{2}{b}\left[v^{j}_{0}\left(\tilde{\theta}w_{0}-\frac{\Theta^{lt}}{2}v^{l}_{0}\right)+\frac{\Theta^{jt}}{2}w_{0}^2 -\frac{1}{2}(\Theta^{lj}+\Theta^{jl})w_{0}v^{l}_{0}\right],
\\
a^{it}&=&-\frac{1}{2}\rho_{0}\Theta^{it}
-\frac{2}{b}\left[v^{i}_{0}\left(\tilde{\theta}w_{0}-\frac{\Theta^{lt}}{2}v^{l}_{0}\right)+\frac{\Theta^{it}}{2}w_{0}^2 -\frac{1}{2}(\Theta^{li}+\Theta^{il})w_{0}v^{l}_{0}\right],
\\
a^{ij}&=&\tilde{\theta}\rho_{0}\delta^{ij}-\frac{\rho_{0}}{2}(\Theta^{ij}+\Theta^{ji})
-\frac{2}{b}\left(\tilde{\theta}{v}^{i}_{0}{v}^{j}_{0}+\frac{\Theta^{it}}{2}w_{0}v^{j}_{0}+\frac{\Theta^{jt}}{2}v^{i}_{0}w_{0}\right)
\nonumber\\
&+&\frac{2}{b}\left[\frac{1}{2}(\Theta^{lj}+\Theta^{jl})v^{i}_{0}v^{l}_{0}+\frac{1}{2}(\Theta^{li}+\Theta^{il})v^{j}_{0}v^{l}_{0}\right].
\end{eqnarray}
In the following we shall define the local sound speed in the fluid and velocity of flow as $c^2_{s}=\frac{b\rho_{0}}{2w^2_{0}}$ and $v^{i}=\frac{v^i_{0}}{w_{0}}$.
The equation (\ref{eqwave}) can be seen as the Klein-Gordon equation in a curved spacetime and can be written as~\cite{Unruh}
\begin{equation}
\label{eqkg}
\frac{1}{\sqrt{-g}}\partial_{\mu}\sqrt{-g}g^{\mu\nu}\partial_{\nu}S_{1}=0,
\end{equation}
being the metric components given in the form
\begin{equation}
\sqrt{-g}g^{\mu\nu}\equiv \frac{b\rho_{0}}{2c_{s}^2}
\left[\begin{array}{clcl}
g^{tt} &\vdots & g^{tj}\\
\cdots\cdots\ &\cdot & \cdots\cdots\\
g^{it} &\vdots & g^{ij}
\end{array}\right],
\end{equation}
where
\begin{eqnarray}
g^{tt}&=&-\tilde{\theta}c_{s}^2
-\left(\tilde{\theta}-\Theta^{jt}v^{j}\right),
\nonumber\\
g^{tj}&=&-\frac{\Theta^{jt}}{2}c^{2}_{s}-\left[\tilde{\theta}v^{j}-\frac{\Theta^{lt}}{2}v^{l}v^{j}+\frac{\Theta^{jt}}{2} -\frac{\Theta^{lj}}{2}v^{l}-\frac{\Theta^{jl}}{2}v^{l}\right],
\nonumber\\
g^{it}&=&-\frac{\Theta^{it}}{2}c^{2}_{s}-\left[\tilde{\theta}{v}^{i}-\frac{\Theta^{lt}}{2}v^{l}v^{i}+\frac{\Theta^{it}}{2}
-\frac{\Theta^{li}}{2}v^{l}-\frac{\Theta^{il}}{2}v^{l}\right],
\nonumber\\
g^{ij}&=&\left[\tilde{\theta}\delta^{ij}-\frac{1}{2}(\Theta^{ij}+\Theta^{ji})\right]c^{2}_{s}
-\left[\tilde{\theta}{v}^{i}{v}^{j}+\frac{\Theta^{it}}{2}v^{j}+\frac{\Theta^{jt}}{2}v^{i}\right]
\nonumber\\
&+&\left[\frac{1}{2}(\Theta^{lj}+\Theta^{jl})v^{i}v^{l}+\frac{1}{2}(\Theta^{li}+\Theta^{il})v^{j}v^{l}\right].
\end{eqnarray} 
In terms of the inverse of $g^{\mu\nu}$ we have the metric of a noncommutative acoustic black hole 
\begin{equation}
\label{invs_g}
g_{\mu\nu}\equiv\frac{\frac{b\rho_{0}}{2c_{s}}}{\sqrt{f}}
\left[\begin{array}{clcl}
g_{tt}&\vdots & g_{ti}\\
\cdots\cdots &\cdot & \cdots\cdots\\
g_{jt}&\vdots & g_{ij}
\end{array}\right],
\end{equation}
where
\begin{eqnarray}
g_{tt}&=&-\left[(\tilde{\theta}-\Theta^{jj})c^{2}_{s}
-\tilde{\theta}{v}^{2}+2\Theta^{jl}v^{j}v^{l}-\Theta^{jt}v^{j}\right],
\nonumber\\
g_{tj}&=&-\frac{\Theta^{jt}}{2}c^{2}_{s}-\left[\tilde{\theta}v^{j}-\frac{\Theta^{lt}}{2}v^{l}v^{j}+\frac{\Theta^{jt}}{2} -\frac{\Theta^{lj}}{2}v^{l}-\frac{\Theta^{jl}}{2}v^{l}\right],
\nonumber\\
g_{it}&=&-\frac{\Theta^{it}}{2}c^{2}_{s}-\left[\tilde{\theta}{v}^{i}-\frac{\Theta^{lt}}{2}v^{l}v^{i}+\frac{\Theta^{it}}{2}
-\frac{\Theta^{li}}{2}v^{l}-\frac{\Theta^{il}}{2}v^{l}\right],
\nonumber\\
g_{ij}&=&[\tilde{\theta}(1+c_{s}^2)-\tilde{\theta}v^2-\Theta^{lt}v^{l}]\delta^{ij}+\tilde{\theta}v^{i}v^{j}.
\end{eqnarray}  
Here $\Theta^{jt}=\theta^{ij}{F_{i}}^{t}=-\theta^{ij}F^{it}=\theta^{ij}E^{i}$ and $\Theta^{jl}=\theta^{ij}{F_{i}}^{l}=-\theta^{ij}F^{il}$. Thus, we find the following components 
\begin{eqnarray}
g_{tt}&=&-[(1-3\vec{\theta}\cdot\vec{B})c^{2}_{s}-(1+3\vec{\theta}\cdot\vec{B})v^2
+2(\vec{\theta}\cdot\vec{v})(\vec{B}\cdot\vec{v})-(\vec{\theta}\times\vec{E})\cdot\vec{v}],
\nonumber\\
g_{tj}&=&-\frac{1}{2}(\vec{\theta}\times\vec{E})^{j}(c^{2}_{s}+1)-\left[2(1+2\vec{\theta}\cdot\vec{B})
-(\vec{\theta}\times\vec{E})\cdot\vec{v}\right]\frac{v^j}{2}+\frac{B^j}{2}(\vec{\theta}\cdot\vec{v})+\frac{\theta^j}{2}(\vec{B}\cdot\vec{v}),
\nonumber\\
g_{it}&=&-\frac{1}{2}(\vec{\theta}\times\vec{E})^{i}(c^{2}_{s}+1)-\left[2(1+2\vec{\theta}\cdot\vec{B})-(\vec{\theta}\times\vec{E})\cdot\vec{v}\right]\frac{v^i}{2}
+\frac{B^i}{2}(\vec{\theta}\cdot\vec{v})+\frac{\theta^i}{2}(\vec{B}\cdot\vec{v}),
\nonumber\\
g_{ij}&=&[(1+\vec{\theta}\cdot\vec{B})(1+c^2_{s})-(1+\vec{\theta}\cdot\vec{B})v^2
-(\vec{\theta}\times\vec{E})\cdot\vec{v}]\delta^{ij}+(1+\vec{\theta}\cdot\vec{B})v^{i}v^{j}.
\nonumber\\
f&=&[(1-2\vec{\theta}\cdot\vec{B})(1+c^2_{s})-(1+4\vec{\theta}\cdot\vec{B})v^2]
-3(\vec{\theta}\times\vec{E})\cdot\vec{v}+2(\vec{B}\cdot\vec{v})(\vec{\theta}\cdot\vec{v}).
\end{eqnarray}
The metric depends simply on the density $\rho_0$, the local sound speed in the fluid $c_s$, the velocity of flow $\vec{v}$, the noncommutativity parameter $\vec{\theta}$ and gauge field components  $\vec{E}, \vec{B}$. 
Notice that the sound speed $c_{s}$ is a function of the electromagnetic field $A_{t}$.
The acoustic line element can be written as
\begin{eqnarray}
\label{metrica}
ds^2&=&\frac{b\rho_{0}}{2c_{s}\sqrt{f}}
\left[g_{tt}dt^2+g_{it}dx^idt+g_{jt}dtdx^j+g_{ij}d{x}^idx^j\right],
\nonumber\\
&=&\frac{b\rho_{0}}{2c_{s}\sqrt{f}}[-{\cal F}(v)dt^2-\vec{\xi}(v)\cdot d\vec{x}dt+ \Lambda(v)dx^2+(1+\vec{\theta}\cdot\vec{B})(\vec{v}\cdot d\vec{x})^2],
\end{eqnarray}
where
\begin{eqnarray}
{\cal F}(v)&=&(1-3\vec{\theta}\cdot\vec{B})c^{2}_{s}-(1+3\vec{\theta}\cdot\vec{B})v^2-(\vec{\theta}\times\vec{E})\cdot\vec{v}
+2(\vec{\theta}\cdot\vec{v})(\vec{B}\cdot\vec{v}),
\\
\Lambda(v)&=&(1+\vec{\theta}\cdot\vec{B})(1+c^2_{s}-v^2)-(\vec{\theta}\times\vec{E})\cdot\vec{v},
\\
\vec{\xi}(v)&=&[2(1+2\vec{\theta}\cdot\vec{B})-(\vec{\theta}\times\vec{E})\cdot\vec{v}]\vec{v}+(1+c^2_s)(\vec{\theta}\times\vec{E})
-(\vec{B}\cdot\vec{v})\vec{\theta}-(\vec{\theta}\cdot\vec{v})\vec{B}.
\end{eqnarray}
Now changing the time coordinate as, 
\begin{eqnarray}
d\tau=dt+\frac{\vec{\xi}(v)\cdot d\vec{x}}
{2{\cal F}(v)},
\end{eqnarray}
we find the acoustic metric in the stationary form
\begin{eqnarray}
ds^2=\frac{b\rho_{0}}{2c_{s}\sqrt{f}}\left[-{\cal F}(v)d\tau^2
+\Lambda\left(\frac{v^iv^j\Gamma+\Sigma^{ij}}{\Lambda{\cal F}(v)}+\delta^{ij}\right)dx^idx^j\right],
\end{eqnarray}
where,
\begin{eqnarray}
\Gamma(v)&=&1+4\vec{\theta}\cdot\vec{B}+(1-2\vec{\theta}\cdot\vec{B})c^2_{s}-(1+4\vec{\theta}\cdot\vec{B})v^2
-2(\vec{\theta}\times\vec{E})\cdot\vec{v}+2(\vec{\theta}\cdot\vec{v})(\vec{B}\cdot\vec{v}),
\\
\Sigma^{ij}(v)&=&[(1+c^2_{s})(\vec{\theta}\times\vec{E})^{i}-(\vec{B}\cdot\vec{v})\theta^{i}-(\vec{\theta}\cdot\vec{v})B^{i}]v^{j}.
\end{eqnarray}
\section{The dispersion relation}
\label{dp}
The sound waves are usually governed by an effective Lorentzian spacetime geometry. In order to study the effect of the spacetime noncommutativity
in such structure we should investigate the dispersion relation. So let us now discuss the dispersion relation.

We now derive the dispersion relation from Eq.(\ref{eqwave}). Since the field $S_{1}$ is real we use the notation
\begin{equation}
S_{1}\sim Re[e^{(i\omega t-i\vec{k}\cdot\vec{x})}], \quad \omega=\frac{\partial S_1}{\partial t}, \quad \vec{k}=\vec{\nabla} S_{1}.
\end{equation}
In this case, the Klein-Gordon equation (\ref{eqwave}) in terms of momenta and frequency, becomes
\begin{equation}
a\omega^2+\sigma\omega+d=0,
\end{equation}
where
\begin{eqnarray}
&&a=(1+\vec{\theta}\cdot\vec{B})(1+c^2_{s})-(\vec{\theta}\times\vec{E})\cdot\vec{v},
\\
&&\sigma=[2(1+2\vec{\theta}\cdot\vec{B})-(\vec{\theta}\times\vec{E})\cdot\vec{v}](\vec{v}\cdot\vec{k})
+(c^2_{s}+1)(\vec{\theta}\times\vec{E})\cdot\vec{k}-(\vec{\theta}\cdot\vec{v})(\vec{B}\cdot\vec{k})-(\vec{B}\cdot\vec{v})(\vec{\theta}\cdot\vec{k}),
\\
&&d=-\{\left[(1+2\vec{\theta}\cdot\vec{B})k^2-(\vec{\theta}\cdot\vec{k})(\vec{B}\cdot\vec{k})\right]c^2_{s}
-(1+3\vec{\theta}\cdot\vec{B})v^2k^2-[(\vec{\theta}\times\vec{E})\cdot\vec{k}](\vec{v}\cdot\vec{k})
\nonumber\\
&&+(\vec{\theta}\cdot\vec{v})(\vec{B}\cdot\vec{k})(\vec{v}\cdot\vec{k})
+(\vec{B}\cdot\vec{v})(\vec{\theta}\cdot\vec{k})(\vec{v}\cdot\vec{k})\}.
\end{eqnarray}
Here we choose $k^{i}=k\delta^{i1}$ ($i=1$, $2$, $3$). Thus, the dispersion relation can be easily found and reads
\begin{eqnarray}
\omega=\frac{-\sigma\pm\sqrt{\sigma^2-4ad}}{2a},
\end{eqnarray}
where
\begin{eqnarray}
\Delta=\sigma^2-4ad&=&4k^2c^2_{s}\left\{ 
\left[(1+3\vec{\theta}\cdot\vec{B})(1+c^2_{s})-(1+4\vec{\theta}\cdot\vec{B})v^2_{1}
-(\vec{\theta}\times\vec{E})\cdot\vec{v}\right]\right.
\nonumber\\
&-&\left.\theta_{1}B_{1}(1+c^2_{s})+(\vec{B}\cdot\vec{v})\theta_{1}v_{1} + (\vec{\theta}\cdot\vec{v})B_{1}v_{1}\right\}.
\end{eqnarray}
We can simplify our result using the following projections $\vec{\theta}\cdot\vec{B}=\theta_3B_3$ and $\vec{\theta}\times\vec{E}=0$, or even for a pure magnetic background, i.e.,  $E=0$, so that
\begin{eqnarray}
\omega=\frac{-(1+2\theta_{3}B_{3})(v_{1}k)
\pm c_{s}k\sqrt{(1+3\theta_{3}B_{3})(1+c^2_{s})-(1+4\theta_{3}B_{3})v^2_1}}{(1+\theta_{3}B_{3})(1+c^2_s)},
\end{eqnarray}
and for $\theta=0$, we recover the result obtained in \cite{Xian}. In the limit $c^2_{s}\ll1$, $v^2_1\ll1$ and for small $\theta_{3}B_{3}$  the dispersion relation is simply given by
\begin{equation}
\omega\approx \pm\frac{c_{s}\sqrt{(1+3\theta_{3}B_{3})}}{(1+\theta_{3}B_{3})}k=\pm c_{s}(1+\frac{1}{2}\theta_{3}B_{3})k.
\end{equation}
This means the group velocity that measures the maximal attainable velocity of a particle in the medium is given by 
\begin{equation}\label{eq-vg}
v_g=\left|\frac{d\omega}{dk}\right|= c_{s}(1+\frac{1}{2}\theta_{3}B_{3}),
\end{equation}
or in terms of deviations in relation to the sound speed $c_s$, the maximal attainable velocity in the medium, then we have 
\begin{equation}
\frac{v_g-c_s}{c_s}=\frac{1}{2}\theta_{3}B_{3}.
\end{equation}
This deviation also appears in recent scenarios with Lorentz violating parameters \cite{ABP,ABP2,codata,Casana:2011bv} with magnitude around $|\beta|\sim|\theta_{3}B_{3}|\sim10^{-6}$, a bound found in BEC physics \cite{codata,Casana:2011bv} --- see also \cite{alfaro}. Notice this implies a {\it supersonic} behavior of a particle with maximal attainable velocity $v_g$. This is pretty similar to the {\it superluminal} behavior recently found in neutrino physics in OPERA experiment \cite{:2011zb} --- see also \cite{OPERA2} for some very recent theoretical issues.
\begin{equation}
\frac{v_\mu-c}{c}\sim 10^{-5},
\end{equation}
where $v_\mu$ is the muon neutrino velocity and $c$ is the light speed. We shall return to this point shortly.

\section{Canonical acoustic black holes}
\label{acoustic-BH-3}
In this section, we shall address the issue of Hawking temperature in the regime of low velocities for the previous study with further
details. Now we consider an incompressible fluid with spherical symmetry. In this case the density $\rho$ is a position independent quantity
and the continuity equation implies that $v\sim\frac{1}{r^2}$. The sound speed is also a constant.

The noncommutative acoustic metric can be written as a Schwarzschild metric type, up to an irrelevant
position-independent factor, as follows,
\begin{eqnarray}
ds^2&=&-\tilde{{\cal F}}(v_{r})d\tau^2+\frac{[v_{r}^2\Gamma+\Sigma+\tilde{{\cal F}}(v_{r})\Lambda]}{\tilde{\cal F}(v_{r})}dr^2
+\frac{(1+c^2_{s})r^2(d{\vartheta}^2+\sin^2\vartheta d\phi^2)}{\sqrt{f}},
\end{eqnarray}
where $\tilde{{\cal F}}(v_{r})=\frac{{\cal F}(v_{r})}{\sqrt{f}}$. In the limit $c^2_s\ll1$
and $v^2\ll1$, we have
\begin{eqnarray}
\label{hz1}
&&\tilde{\cal F}(v_{r})=\frac{\left[(1-3\vec{\theta}\cdot\vec{B})c^2_{s}-(1+3\vec{\theta}\cdot\vec{B})v^2_{r}-(\vec{\theta}\times\vec{E})_{r}{v}_{r}
+2(\theta_{r}B_{r}v^2_{r})\right]}{\sqrt{(1-2\vec{\theta}\cdot\vec{B})
-3(\vec{\theta}\times\vec{E})_{r}{v}_{r}}}.
\end{eqnarray}
Now using the relation $v_{r}=c_{s}\frac{r^2_{h}}{r^2}$ in the equation above, where $r_{h}$ is the event horizon, the radius at
which the flow speed exceeds the sound speed in the fluid, we have
\begin{eqnarray}
\tilde{\cal F}(r)=\frac{c^2_{s}\left[(1-3\vec{\theta}\cdot\vec{B})-(1+3\vec{\theta}\cdot\vec{B}-2\theta_{r}B_{r})\frac{r^4_{h}}{r^4}
-(\vec{\theta}\times\vec{E})_{r}\frac{r^2_{h}}{c_{s}r^2}\right]}
{\sqrt{(1-2\vec{\theta}\cdot\vec{B})
-3[(\vec{\theta}\times\vec{E})_{r}]\frac{r^2_{h}}{c_sr^2}}}.
\end{eqnarray}
In this case the Hawking temperature is given by
\begin{equation} 
T_{H}=\frac{\tilde{\cal F}^{\prime}(r_h)}{4\pi}=c^{2}_{s}\frac{[1+3\vec{\theta}\cdot\vec{B}-2\theta_{r}B_{r}+(\vec{\theta}\times\vec{E})_{r}/2c_{s}]}
{\sqrt{1-2\vec{\theta}\cdot\vec{B}-3(\vec{\theta}\times\vec{E})_{r}/c_{s}}}
(\pi r_{h})^{-1}+O(\theta^2),
\end{equation}
that for $\theta_r=0$, $\vec{\theta}\cdot\vec{B}=\theta_3B_3$,  $\vec{\theta}\times\vec{E}=0$ (or  $E=0$)  with small $\theta_3B_3$,
\begin{equation} 
T_{H}=c^{2}_{s}\frac{(1+3\theta_3B_3)}
{\sqrt{1-2\theta_3B_3}}
(\pi r_{h})^{-1}=\left(1+4\theta_3B_3\right)\frac{c^2_{s}}{(\pi r_{h})}.
\end{equation}
For $\theta=0$ the usual result is obtained, otherwise one can see from (\ref{eq-vg}) that the Hawking temperature goes like $T_H\to v_g^8T_H$, for $\theta_3B_3$ small enough.

\begin{figure}
\includegraphics[width=6.0cm]{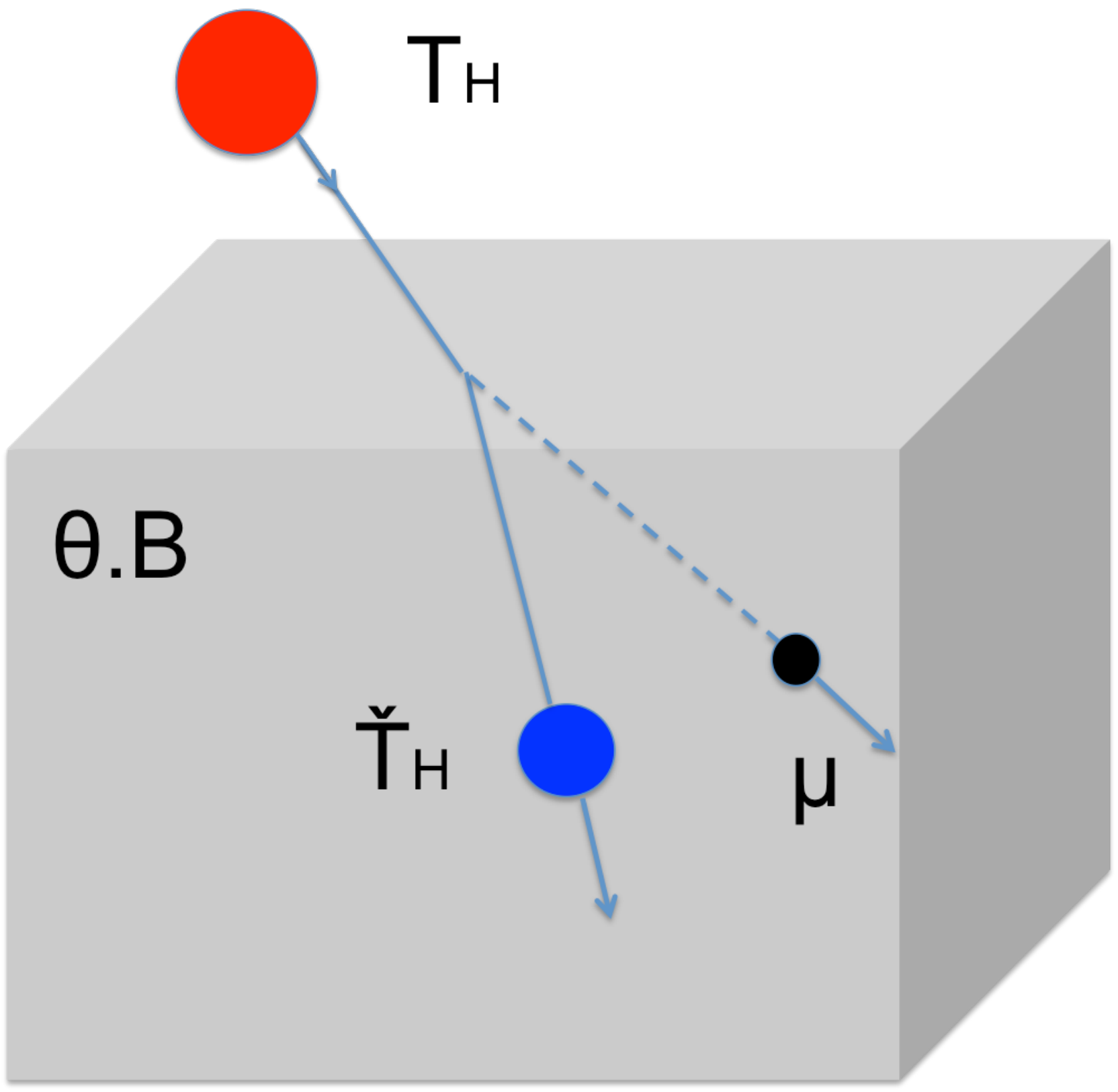}
\caption{The black hole point of view of muon neutrino production through a change $\theta_3B_3$ of the vacuum due to the spacetime noncommutativity. The Hawking temperature variation $\Delta T_H=T'_H-T_H$ is related to the black hole decay into muon neutrino that follows its way in the changed medium with velocity faster than light speed.}
\label{fig1}
\end{figure}

Let us now analyze this result more carefully. The formula above can be rewritten in terms of variations of the Hawking temperature with respect to the {\it changing of medium} due to 
spacetime noncommutativity and `strong' magnetic field component as follows 
\begin{equation} \label{eq-5-1}
\frac{\Delta T_H}{T_H}=4\theta_3B_3.
\end{equation}
This allows us to write the formula 
\begin{equation}\label{deltaT}
\frac{v_g-c_s}{c_s}=\frac{1}{2}\theta_{3}B_{3}=\frac{1}{8}\frac{\Delta T_H}{T_H}.
\end{equation}
As we mentioned earlier the Hawking radiation was shown by Unruh to be a purely kinematic effect of quantum field theory, so we can study the Hawking radiation process in completely different physical systems. Thus, acoustic horizons possess many of the properties associated with the event horizons of general relativity. As such we can assume that (\ref{deltaT}) is also valid for general relativity, i.e.,
\begin{equation}\label{deltaTg}
\frac{v-c}{c}=\frac{1}{2}\theta_{3}B_{3}=\frac{1}{8}\frac{\Delta T_H}{T_H}.
\end{equation}
Now we assume that the muon neutrino production is accompanied by a stronger probing of the spacetime noncommutativity that is equivalent to feel a change in the vacuum according to (\ref{eq-5-1}) --- see 
Fig.~\ref{fig1}. We also consider the Hawking radiation temperature (e.g. due to mini black holes with $r_h$ sufficiently small) is about a proton mass, i.e. $T_H\simeq $1GeV and its variation is the energy sufficient to produce the heaviest muon neutrino (expected upper bound), i.e., $\Delta T_H\simeq170$ keV. Plugging these quantities into (\ref{deltaTg}) gives the nice result 
\begin{equation}\label{deltaTg-R}
\frac{v-c}{c}=2.125\times 10^{-5}.
\end{equation}
This is in accord with OPERA experiments \cite{:2011zb}. 

There is also another perspective for the above scenario in order to explain superluminal neutrinos if we consider the Hawking temperature due to {\it surface gravity} of objects such the Earth in the OPERA experiments or a {\it blue supergiant star} in the case of supernova SN1987A measurements or even in terms of a {\it surface wave} related to the OPERA superluminal muon neutrino beam. It is easy to recast the formula (\ref{deltaT}) in terms of surface gravity, i.e., $T_H\sim g$ where $g$ is the surface gravity of an object.  The surface gravity of the Earth at some small height $h$ above its surface at radius $R_e$ can be written as follows 
\begin{equation}\label{Earth}
g_h=\frac{GM_e}{(R_e+h)^2}\sim\frac{GM_e}{R_e^2}\left(1-\frac{2h}{R_e}\right),
\end{equation}
and the surface gravity fluctuation due to such small height (or deviation) is given by
\begin{equation}\label{Earth2}
\Delta g_h\equiv g-g_h=\frac{2GM_e}{R_e^3}h,
\end{equation}
where $g$ is the surface gravity on the Earth surface. Now we can write 
\begin{equation}\label{Earth3}
\frac{\Delta T_H}{T_H}\equiv\frac{\Delta g_h}{g_h}\sim\left(\frac{2h}{R_e}-\frac{4h^2}{R_e^2}\right).
\end{equation}
The last term contributes to subluminal velocities, but here it is highly suppressed and we shall consider below just the leading term. Thus, let us now rewrite (\ref{deltaT}) as follows
\begin{equation}\label{Earth4}
\frac{v-c}{c}=\frac18\frac{\Delta g_h}{g_h}\sim\frac14\frac{h}{R_e}.
\end{equation}
Since the Earth radius $R_e\sim 10^6$ m, then plugging the OPERA result $\frac{v-c}{c}\sim 10^{-5}$ into (\ref{Earth4}) one finds $h\sim40$ m. Now applying this formula for a  blue supergiant star, that is, $R_e\to R_{\rm star}\sim 40\times R_{\rm sun}$ and the fact that $R_{\rm sun}\sim 100 R_e$ one finds 
\begin{equation}\label{SN1987A}
\frac{v-c}{c}\sim\frac14\frac{h}{R_{\rm star}}=\frac14\frac{40{\rm m}}{40\times 10^8{\rm m} }\sim 10^{-9},
\end{equation}
which agrees with supernova SN1987A measurements. In this perspective fluctuations of surface gravity in both the Earth and blue supergiant stars produce a virtual particle with mass $m_h\sim \frac{1}{h}\sim 10^{-8}ev$ that affects the muon neutrino velocities. This goes in the same line of the recent investigations by considering background influence in the neutrino superluminality \cite{OPERA2}.

The noncommutative acoustic metric can also be written in a Kerr-like form. We now address the issue of rotating black holes by using the projections above such that we can rewrite Eq.(\ref{metrica}) as
\begin{eqnarray}
ds^2&=&\frac{b\rho_{0}}{2c_{s}\sqrt{f}}\left[-[(1-3\theta_{z}B_{z})c^{2}_{s}-(1+3\theta_{z}B_{z})(v^2_{r}+v^2_{\phi})]dt^2
-2(1+2\theta_{z}B_{z})v_{r}drdt\right.
\nonumber\\
&-&\left.2(1+2\theta_{z}B_{z})v_{\phi}rd{\phi}dt
+(1+\theta_{z}B_{z})[v^2_{r}dr^2+r^2v^{2}_{\phi}d\phi^2+2v_{r}v_{\phi}rdrd\phi]\right.
\nonumber\\
&+&\left.[(1+\theta_{z}B_{z})(1+c^2_{s}-v_{r}^2-v^2_{\phi})](dr^2+r^2d\phi^2+dz^2)\right].
\end{eqnarray}
However, exploring the original solutions as spherically symmetric solutions with $v_{z}=0$, $v_{r}\neq0$ and v$_{\phi}\neq 0$ one can show
that they can be written in a Kerr-like form
\begin{eqnarray}
\label{Kerr}
ds^2&=&\frac{b\rho_{0}}{2c_{s}}\left[-N^{2}d\tau^2+M^2dr^2+Q^{2}r^2d\varphi^2+Z^2dz^2
+\frac{(1+\theta_{z}B_{z})\left[(1+\theta_{z}B_{z})v_{\phi}d\tau-rd\varphi\right]^2}{\sqrt{f}}\right]
\end{eqnarray}
where we have the Kerr-like components as
\begin{eqnarray}
&&N^2=\frac{(1-3\theta_{z}B_{z})c^{2}_{s}-(1+3\theta_{z}B_{z})v^2_{r}}{\sqrt{f}},
\quad \quad M^2=\frac{{\cal{F}}c^2_{s}}{[(1-3\theta_{z}B_{z})c^{2}_{s}-(1+3\theta_{z}B_{z})v^2_{r}]\sqrt{f}},
\\
&&Q^2=\frac{(1+\theta_{z}B_{z})(c^{2}_{s}-v^2_{r})}{\sqrt{f}},
\quad\quad\quad\quad\quad\quad\quad Z^2=\frac{(1+\theta_{z}B_{z})(1+c_{s}^2-v^2)}{\sqrt{f}},
\\
&&{\cal{F}}=(1-2\theta_{z}B_{z})(1+c_{s}^2)-(1+4\theta_{z}B_{z})v^2
+\frac{6\theta_{z}B_{z}v^2_{r}v^2_{\phi}}{c^2_s-v^2_r},
\\
&&f=(1-2\theta_{z}B_{z})(1+c_{s}^2)-(1+4\theta_{z}B_{z})v^2
\end{eqnarray}
and the coordinate transformations we have used are
\begin{eqnarray}
&&d\tau=dt+\frac{(1+2\theta_{z}B_{z})v_{r}dr}
{[(1-3\theta_{z}B_{z})c_{s}^2-(1+3\theta_{z}B_{z})v_{r}^2]},
\nonumber\\
&&d\varphi=d\phi+\frac{v_{r}v_{\phi}dr}{r[c_{s}^2-v_{r}^2]}.
\end{eqnarray}
Now we find the important components
\begin{eqnarray}
g_{\tau\tau}=\frac{-(1-3\theta_zB_z)c_s^2+(1+3\theta_zB_z)(v_r^2+v_\phi^2)}{\sqrt{f}}, \qquad g_{rr}=\frac{{\cal{F}}c^2_{s}}{[(1-3\theta_{z}B_{z})c^{2}_{s}-(1+3\theta_{z}B_{z})v^2_{r}]\sqrt{f}},
\end{eqnarray}
where we have made the approximation $(1-\theta_zB_z)^3\simeq(1-3\theta_zB_z)$ which is valid for $\theta_zB_z$ sufficiently small.
For a planar solution (assuming $z=0$) the velocities assume the form  $v_r=\frac{A}{r}$ and $v_\phi=\frac{B}{r}$. After substituting this into equations above we are able to find the ergosphere radius and the horizon via coordinate singularity through the following equations $g_{\tau\tau}(r_e)=0$ and $g_{rr}(r_h$)=0, respectively. The corresponding radii read 
\begin{eqnarray}
r_e=(1+3\theta_3B_3)\frac{(A^2+B^2)^{1/2}}{c_s}, \qquad r_h=(1+3\theta_3B_3)\frac{|A|}{c_s}
\end{eqnarray}
Notice that $3\theta_zB_z$ stands for the Lorentz violating parameter $\beta$ in our previous results \cite{ABP,ABP2} for Lorentz violating acoustic black holes. Starting from this point all analysis made in \cite{ABP,ABP2} applies here. Many interesting studies can be followed from this point. One of them is the ÔsuperresonanceÕ \cite{Basak:2002aw,ABP2,Zhang:2011zzh} which is an analog of the superradiance phenomenon in gravitational black holes, but a detailed study on this subject is out of the scope of this paper. We shall consider this study in a forthcoming publication.

\section{Conclusions}
\label{conclu}

One of the main results of the present paper is that supersonic and analogously superluminal particles can be understood in terms of $\Delta T_H/T_H$, where $T_H$ is the Hawking temperature.
We have considered this study in noncommutative acoustic black holes in a noncommutative  Abelian Higgs model. The Abelian Higgs model is good to describe high energy physics
and noncommutative Abelian Higgs model can also describe Lorentz symmetry violation in particle physics in high energy. Thus our results suggest that in addition to the expected gravitational mini black holes formed in high energy experiments one can also expect the formation of acoustic black holes together. 

The model also develops several similarities with respect to Lorentz violating acoustic black holes studied  in \cite{ABP,ABP2}. One of the consequences is that the acoustic Hawking temperature is changed such that it depends on the group speed which means that, analogously to the gravitational case \cite{syb,adam}, the Hawking temperature is {\it not} universal for all species of particles. It depends on the maximal attainable velocity of this species. In the context of gravitational black holes this has been previously studied and appointed as a sign of possibly violation of the second law of the thermodynamics.  Furthermore, the acoustic black hole metric in our model can be identified with 
an acoustic Kerr-like black hole. As we explicitly have shown in  \cite{ABP2}, using a similar Lorentz violating setup, the spacetime noncommutativity should also affect the rate of loss of mass (energy) of the black hole. 
Thus for suitable values of noncommutative parameter a wider or lower spectrum of particle wave function can be scattered with increased amplitude by the acoustic black hole. The  superressonance
and its increasing/decreasing phenomenon have been previously studied in \cite{Basak:2002aw,ABP2,Zhang:2011zzh,SBP}. 

\acknowledgments

We would like to thank CNPq, CAPES, PNPD/PROCAD -
CAPES for partial financial support.

\end{document}